\documentclass[usenatbib]{mn2e}
\input epsf

\title[Millisecond dips of Sco X-1 and TNO size distribution]{Millisecond 
dip events in the 2007 RXTE/PCA data of Sco X-1 and the TNO size distribution}
\author[Liu, Chang, Liang, and King]{Chih-Yuan Liu$^1$, Hsiang-Kuang Chang$^{1,2}$\thanks{E-mail:
hkchang@phys.nthu.edu.tw}, Jau-Shian Liang$^1$, and Sun-Kun King$^3$ 
\\
$^{1}$Department of Physics, National Tsing Hua University, Hsinchu 30013, Taiwan\\ 
$^{2}$Institute of Astronomy, National Tsing Hua University, Hsinchu 30013, Taiwan\\ 
$^{3}$Institute of Astronomy and Astrophysics, Academia Sinica, Taipei 10617, Taiwan} 

\begin{document}

\date{January 2008; April 2008}

\pagerange{\pageref{firstpage}--\pageref{lastpage}} \pubyear{2008}

\maketitle

\label{firstpage}

\begin{abstract}
Millisecond dips in the RXTE/PCA archival data of Sco X-1 taken from 1996 to 2002 were reported recently.
Those dips were found to be most likely caused by instrumental dead time but may also contain some 
true astronomical events, which were interpreted as
the occultation of X-rays from Sco X-1
by Trans-Neptunian Objects (TNO) of 100-m size.
Here we report the results of search for millisecond dip events 
with the new RXTE/PCA data of Sco X-1 taken in year 2007. 
Adopting the same selection criteria as that in the previous study, 
we found only 3 dip events in 72-ks data, 
much fewer than the 107 events found in the 560-ks data taken from 1996 to 2002 reported earlier. 
The new data provides more detailed information of individual `very large events' (VLEs), 
which is not available in the
old archival data. 
Although the number of VLEs does not obviously increase during the occurrence of dip events, 
all the 3 dip events are coincident in time with VLEs that have no flags set for any 
of the propane or the 6 main xenon anodes. 
It is a strong indication of instrumental effects.
No significant dips which might be real occultation by
60 -- 100 m TNOs were observed. 
With only 72-ks data, however, the previously proposed possibility that about 10 percent 
of the dip events might not be instrumental still cannot be strictly excluded. 
Using the absence of those anomalous VLEs as the criterion for identifying 
non-instrumental dip events, we found,
at a lower confidence level, 4 dip events of duration 8 - 10 ms in the 72-ks data. 
Upper limits to the size distribution of TNOs at the small size end are suggested.
\end{abstract}


\section{Introduction}

The outer solar system bodies in the Edgeworth-Kuiper Belt and in the Oort Cloud
are relics of the protostellar disk of the sun. 
These objects in the trans-Neptunian region witnessed 
the dynamical and collisional history of the early solar system.
Their properties, such as the total mass, size distribution, and orbital parameters, 
are therefore valuable knowledge to the understanding of the formation of our planetary system.

The first Trans-Neptunian Object (TNO) other than Pluto was discovered in 1992 
\citep{jewitt93}. More than 1000 TNOs have been found since then.
These are all TNOs larger than 100 km. Smaller ones are too dim to detect. 
With much effort in data processing by integrating images over all possible TNO orbits,
a Hubble Space Telescope survey reported the detections 
of 3 TNOs of size about 30 km 
\citep{bernstein04}. 
For even smaller ones, occultation of background 
stars as a way to study their properties was proposed 
thirty years ago 
\citep{bailey76,brown97,roques00,cooray03}.
Searches for such occultation events in optical bands have 
been conducted by some groups but without 
any definite detection so far. 
Some possible detections of TNOs of size about 300 m with 20-Hz photometry using
the Bernard Lyot 2-m telescope of the Pic du Midi Observatory and the 4.2-m William Herschel Telescope
at La Palma
were reported 
\citep{roques03,roques06}.

At a shorter wavelength, millisecond dip events in the RXTE/PCA light curve of Sco X-1
were discovered
\citep{chang06}, which were attributed to occultation caused by small TNOs of 100-m size.
Later, signatures of possible instrumental effects related to high-energy cosmic rays
for many of these dip events were found 
\citep{jones06} and their nature 
became uncertain. 
A further analysis of 107 dip events in 560-ks RXTE archival data 
spanning over 7 years from 1996 to 2002
indicated that the evidence for instrumental effects was not conclusive, 
mainly because of the coarse time
resolution of the relevant house keeping data, 
and the possibility that about 10\% of these dip events are truly astronomical cannot be ruled out
\citep{chang07,jones07}.

To distinguish instrumental and astronomical dip events in the RXTE/PCA light curve of Sco X-1 
and to explore the origin of the previously unkown instrumental effect, 
new RXTE observations of Sco X-1 with a data mode recording detailed information of individual RXTE/PCA
very large events (VLEs) at a time resolution of 125 $\mu$s have been scheduled.
These observations started from June 2007 with 2 or 3 observations per week until October 2007. 
The observation will resume from February 2008 for one more year. 
Here we report the results of the observations in 2007, which already yielded useful information.
We found that the presence of `anomalous VLEs', explained in the next section, can be used
as an indicator for instrumental effects. 
In the 72-ks 2007 RXTE/PCA data, 3 significant dip events similar to
those reported earlier were found and they are all instrumental.
At a less significant level of
5-$\sigma$, 21 dip events were found and 17 of them are instrumental.
With the 4 possible non-instrumental dip events,
which are not inconsistent with random fluctuation, upper limits to the TNO size distribution
at 100 - 300 m are suggeted.  
These results are helpful for the analysis of new RXTE/PCA data in the future 
and for the design and operation of other X-ray space missions which may be
suitable for TNO occultation search, such as India's ASTROSAT 
and the proposed Extreme Physics Explorer 
\citep{elvis06}.  

\vspace{-0.5cm}
\section{Dip events and the anomalous VLEs}

The 2007 RXTE observation of Sco X-1 started from 
June 13 and ended on October 2.
To have a higher count rate, we used only those data taken with 3 or more Proportional Counter Units (PCUs).
Technical details of RXTE instrumentation can be found in the RXTE web site and also in 
\citet{jahoda06}.
A brief description of issues most relevant to our analysis can be found in
\citet{chang07}.
Most of the observations were conducted with 4 or 5 PCUs.
The total amount of the 2007 RXTE/PCA data used in this study is about 72 ks.
To search for millisecond dip events, we adopted exactly the same procedure employed in previous studies 
\citep{chang06,chang07}.
We examined the light curves of Sco X-1 
binned in different bin sizes from 1 ms to 10 ms.
The deviation in photon counts of each bin in the light curve was determined by
comparing its count number with the average counts and the count variance in an 8-s running window.

In our previous studies, 
the significance level adopted for identifying a dip event 
was set at the random probability being $10^{-3}$ 
for the 320-ks data employed in the first study
\citep{chang06}. 
That corresponds to a deviation of $6.5 \sim 6.8\,\, \sigma$, 
depending on the bin size used in the search.
To compare with the previous dip event rate, we used the same deviation level 
as the selection threshold to identify dip events in the new data.
In total, in the 72-ks data, 3 dip events were found.
Their light curves are shown in Fig.\ 1 -- 3 and epochs in Table 1.
\begin{table}
\begin{center}
\begin{tabular}{lll}
\hline
\hline
Event  & Epoch (MJD) & Significance \\
\hline
1 & 54265.453357412 & -8.4 $\sigma$ (2 ms) \\
2 & 54333.916753871 & -11.5 $\sigma$ (8 ms) \\
3 & 54351.054678136 & -6.8 $\sigma$ (2 ms) \\
\hline
A & 54290.842075508 & -5.2 $\sigma$ (9 ms) \\
B & 54292.078448234 & -5.0 $\sigma$ (10 ms) \\
C & 54306.097132898 & -5.0 $\sigma$ (8 ms) \\
D & 54343.145430422 & -5.0 $\sigma$ (9 ms) \\
\hline
\hline
\end{tabular}
\caption{Three most significant dip events and four non-instrumental dip events found in the 72-ks RXTE/PCA
data of Sco X-1 taken in 2007. 
As explained in the main text, Events 1, 2 and 3 are all instrumental and Events A, B, C and D are not
inconsistent with random fluctuations.
The 'Significance' column shows the most significant deviation of the time bins associated with the dip event.
Noted in the parentheses are the corresponding bin size. 
For the exact definition of deviation and the dip search procedure, see
\citet{chang06,chang07}. 
}
\end{center}
\end{table}
\begin{figure}
\epsfxsize=8.4cm
\epsffile{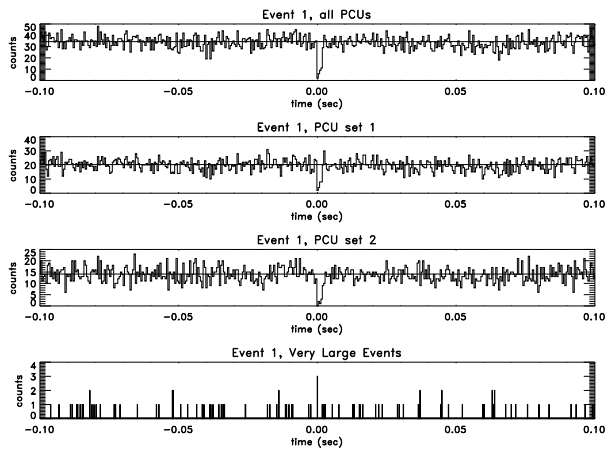}
\caption{RXTE/PCA light curves of Event 1.
Time zero refers to the epoch of Event 1, as listed in Table 1.
The upper three panels are plotted with a 0.5-ms time resolution. 
The light curve in the top panel is the sum of the two in the middle two panels.
Plotted in the bottom panel is the count of 'very large events' (VLEs) at a resolution of 
125 $\mu$s, which is also the designed resolution in the data.
Some VLEs arrive at the same time (within 125 $\mu$s).
At the epoch of the dip events, the VLE count does not obviously increase;
see Fig.\ 2 for a better example. However, all the 3 VLEs at the epoch of Event 1, 
and also those 2 for Event 2 and those 3 for Event 3, are anomalous in the sense that
they do not have any flag set for any anode of the six main xenon layer or the propane layer.
This kind of VLEs are called anomalous VLEs (AVLEs) in this paper and is used as 
the indication of a previously unknown instrumental effect.
}
\end{figure}
\begin{figure}
\epsfxsize=8.4cm
\epsffile{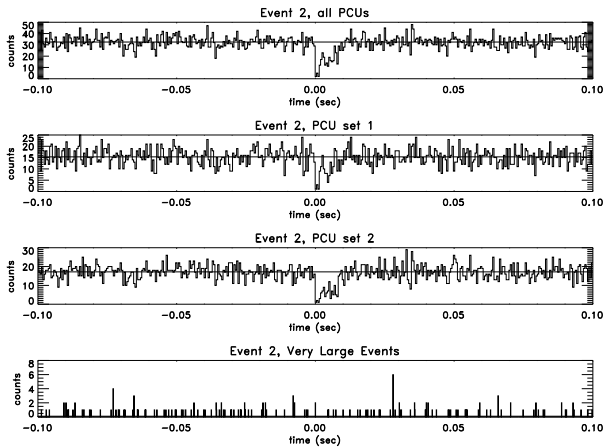}
\caption{RXTE/PCA light curves of Event 2.
See the caption of Fig.\ 1 for more explanations.
We note that at the 28th ms after the Event 2 epoch, 6 VLEs arrive at the same time.
There is nothing unusual in the light curves shown in the upper three panels at that epoch.
The number of VLEs is apparently not related to the occurence of dip events.
}
\end{figure}
\begin{figure}
\epsfxsize=8.4cm
\epsffile{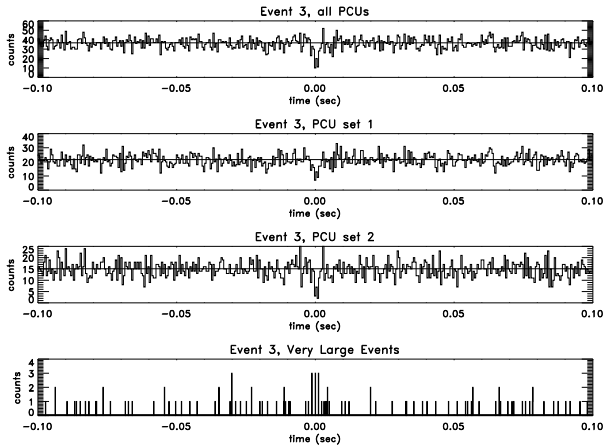}
\caption{RXTE/PCA light curves of Event 3.
See the caption of Fig.\ 1 for more explanations.
}
\end{figure}

The new observation has data modes recording information 
more detailed than before.
In these data modes, detected photons are grouped into 2 sets. 
Set 1 includes those hitting PCU0, PCU2 and PCU4.
Photons detected by PCU1 and PCU3 are in Set 2. 
Light curves of the 3 dip events from these two sets 
and their combination are shown in Fig.\ 1 -- 3.
Furthermore, a newly designed data mode records individual VLEs 
with their detector (PCUs) and anode information 
at a time resolution of 125 $\mu$s.
A VLE is an event that deposits more than about 100 keV energy in any one of
the six active xenon-layer anodes or the propane-layer anode 
in each PCU of the PCA on board RXTE. 
The VLE counts around the epochs of these 3 dip events 
are also plotted in Fig.\ 1 -- 3.

The VLE counts do not particularly increase at the dip epochs.
However, all the 3 dips are coincident in time with 2 or 3 VLEs that
set no flags in the propane or the six main xenon anodes. 
We shall call this kind of VLEs `anomalous VLEs' (AVLEs) hereafter.
It is not yet clear to us 
how AVLEs cause these dips and what AVLEs really are.
Although further investigation on AVLEs requires detailed technical knowledge 
about RXTE/PCA and is beyond our capability, it is very likely that
they are due to high-energy cosmic rays which induce particle showers
in the instrument 
\citep{jones06,jones07}.
This possiblity is further supported by the excess 
of the so-called `remaining count' 
recorded in the RXTE/PCA standard-1 data mode at dip epochs
\citep{jones06,chang07}.
Based on the occurrence of AVLEs at all the 3 dip epochs,
we will use the presence of AVLEs as an indicator of
instrumental effect to look for possible non-instrumental dips. 
We note, however, the presence of AVLEs is 
a sufficient condition to indicate
instrumental effects. Up to this point, one cannot be sure whether
it is also a necessary one.
At the epoch of Event 2, for example, 
the two AVLEs were detected by PCU2 and PCU4,
both included in PCU set 1, 
but the light curve of PCU set 2 also shows the dip.

The number of dips found in the 2007 data seems much smaller than before.
It is 3 in 72 ks, compared with 107 in 560 ks reported earlier 
\citep{chang07}.
We note that, however, in the second half of year 1999, although the amount of the data used is 59 ks,
no such dip events 
(significance higher than $6.5 \sim 6.8\,\, \sigma$) was found. 
The dip event rate apparently varies. 
In the data prior to year 2007, there is no way to identify AVLEs.
Although the 3 dips reported here are all instrumental, 
the possibility that 10\% of the earlier 107 events may be astronomical, 
that is, at a rate of about one per 50 ks,
cannot be strictly excluded by the current non-detection in the 72-ks data.

\vspace{-0.5cm}
\section{Possible non-instrumental dips at a lower significance level}

Equipped with the possibility of identifying non-instrumental dips 
by the absence of AVLEs,
we proceeded to find dip events at a lower significance level. 
We tentatively set the deviation threshold at 5 $\sigma$. 
With this threshold, we found in total 21 dips, 
in addition to the 3 more significant ones described in the last section.
Among the 21 dips, 17 are associated with 1 to 4 AVLEs. 
In the data we used for this study, the average VLE rate is
about 120 counts per second per PCU.
The AVLEs happen at an average rate of 
about 1.8 counts per second per PCU, which is much smaller than the VLE rate.
The association of 17 dips among the 21 with AVLEs 
further supports the validity of using AVLEs as the indication of 
instrumental effects.
We therefore consider these 17 dip events being instrumental.

Light curves of the 4 dip events not associated with any AVLEs 
are shown in Fig.\ 4 and their epochs are listed in Table 1.
\begin{figure}
\epsfxsize=8.4cm
\epsffile{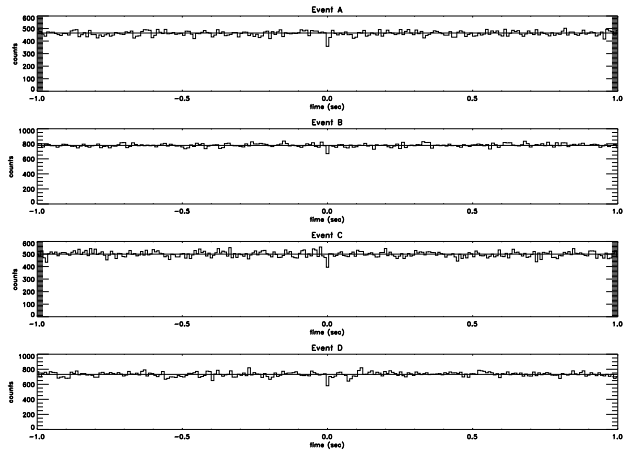}
\caption{RXTE/PCA light curves of the 4 non-instrumental dip events.
Time resolution of these curves is 9, 10, 8 and 9 ms for Event A, B, C and D respectively. 
Time zero refers to the epochs of these events, as listed in Table 1.
}
\end{figure}
These 4 dip events were picked out from searches with bin size of 8 ms (Event C),
9 ms (Events A and D), and 10 ms (Event B) respectively at the 5-$\sigma$ level.
In the 72-ks data, assuming a Gaussian distribution for the random fluctuation,
we expect to see about 2 bins with a negative deviation more than 5-$\sigma$ in these searches.
Although the Gaussian distribution in fact overestimates the number of expected bins at the negative
deviation end because of the Poisson nature of the data, it is actually not much in the current case,
as shown in Fig.\ 5.  
\begin{figure}
\epsfxsize=8.4cm
\epsffile{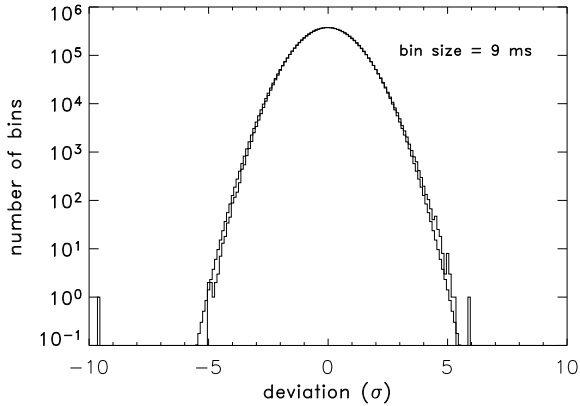}
\caption{The deviation distribution of the Sco X-1 RXTE/PCA light curve with 9-ms bins. 
See \citet{chang06} for the exact definition of the deviation.
The thick histogram is from the 72-ks RXTE/PCA data of Sco X-1 
and the thin one is the Gaussian distribution, plotted for comparison. 
To show the relatively small number of events with large deviations, 
the ordinate is plotted in logarithmic scales. 
The bin at $-9.6\,\,\sigma$ is associated with Event 2, whose light curve is shown in Fig.\ 2.
Those bins at about $-5\,\,\sigma$ are not inconsistent with random fluctuations. 
}
\end{figure}
These 4 dip events are therefore not inconsistent with random fluctuations. 
We will use them to estimate upper limits to the TNO size distribution at the level set by
assuming they are real occultation events.

\section{Upper limits to the TNO size distribution} 

The durations of these 4 events are roughly 8 to 10 ms.
A rule-of-thumb estimate is that they correspond to a shadow of 240 to 300 m, 
assuming they are occultation events. 
On the other hand, if they are caused by TNOs of that size 
(here we ignore all the complexity of diffraction patterns and the crossing location in the shadow), 
those TNOs should be at a distance
much farther than that of most TNOs discovered to date, because their shadows are very shallow.
Let's consider this with the notion of the 'flux-drop fraction' defined in 
\citet{chang07}.
The flux-drop fraction of these 4 events is about 0.2 or so.
From Fig.\ 13 in 
\citet{chang07} one can see that 
at 40 AU a flux drop of 0.3 corresponds to the object size being about 30 m,
that is, one Fresnel scale,
which is $\sqrt{\lambda d/2}$, where $\lambda$
is the wavelength and $d$ is the distance. 
We use $\lambda=0.3$ nm (4 keV) because most of the RXTE/PCA-detected photons from Sco X-1
are at this energy. 
We note that, 
for a given ratio of the object size to the square root of the distance,
the diffraction patterns are the same, when expressed in units of 
the Fresnel scale 
(e.g.\ \citet{roques87}).
The flux-drop
fraction is therefore also the same for these patterns. 
Since the flux drop of the 4 possible non-instrumental events
is about 0.2, instead of 0.3, 
the distances to the objects causing these events
will be beyond 4000 AU, but not by orders of magnitudes, if
 their sizes are all about 300 m. 
On the other hand, their sizes are likely  to be smaller than 300 m.
At first, assuming a shadow-crossing speed of 30 km/s, their duration
implies sizes of 240 - 300 m. Secondly, the shadow-crossing speed is 
probably smaller than 30 km/s, so the size could be even smaller.
To produce a certain level of flux drop, 
a smaller size implies a smaller distance. 
To simplify the discussion, 
we adopt 4000 AU as a representative distance to these objects.
This distance is in the inner Oort Cloud region.

The above estimate is of course a rough one, which we consider adequate at the current stage.
A more detailed approach may be devised by considering the event duration and the flux-drop fraction
as functions of the TNO size and distance, with the RXTE velocity in the solar system at
the epoch of each particular event, assumed orbital parameters of the TNO, assumed TNO shape and
shadow-crossing location (the impact parameter), and the corresponding diffraction
pattern. Such an approach will be reported in details in another forthcoming paper.
Here we contend ourselves with the above estimate.
We propose to derive an upper limits to the total number of a certain population of TNOs
located around 4000 AU at the level of assuming the 4 dip events being real occultation 
by objects of size 200 - 300 m.  
From Eq.(4) in 
\citet{chang07}, which we quote below for convenience,
\begin{equation}
\left(\frac{{\rm d}N}{{\rm d}\log s}\right)_{s_1<s<s_2}
=
\frac{d^2\Omega_{\rm A}}{v(s_2-s_1)}\frac{N}{T}\,\ln 10
\,\,\, ,
\label{szd}
\end{equation}
and with $\Omega_{\rm A}=3.4$,
$v=30$ km/s, 
and $T=72$ ks, we have in the size range from 200 m to 300 m 
\begin{equation}
\left(\frac{{\rm d}N}{{\rm d}\log s}\right)
\approx 5.2\times 10^{19}
\,\,\, .
\end{equation}
This upper limit is plotted in Fig. 6 with the label `(A)'.
In Eq.(\ref{szd}) $\Omega_{\rm A}$ is the sky area covered by the population of TNOs in question.
It may not be equal to 3.4 as derived from the CFHT survey for TNOs larger than 100 km
\citep{trujillo01,chang07}, but the difference will not be orders of magnitudes.
We in fact do not have the knowledge about that for objects in the inner Oort Cloud.

Since in the 72-ks data we did not find any dip events possibly caused by TNOs at about 43 AU,
we proceed to set an upper limit to the size distribution in the range of 100 - 300 m, which 
roughly corresponds to dip durations of 3 - 10 ms. 
A mimic 2-$\sigma$ upper limit can be set at the level of
4 events in 72-ks data. This is of course not in a rigorous statistical sense, but just to  
illustrate the order of magnitude in the size distribution. From Eq.(\ref{szd}), we have  
$\left(\frac{{\rm d}N}{{\rm d}\log s}\right)
\approx 3.0\times 10^{15}$.
This upper limit is plotted in Fig. 6 with the label `(B)'.
In the size range of 60 - 100 m, 
\citet{chang07} suggested possible detection of about 12 events.
However, the 3 significant dips reported in Section 2 
are all instrumental and no dips 
that might be real occultation by 60 - 100 m TNOs were observed. 
We therefore set the upper limit at the level of 12 events in 560 ks
for the size range of 60 - 100 m. 
This upper limit is plotted in Fig. 6 with the label `(C)'.
\begin{figure}
\epsfxsize=8.4cm
\epsffile{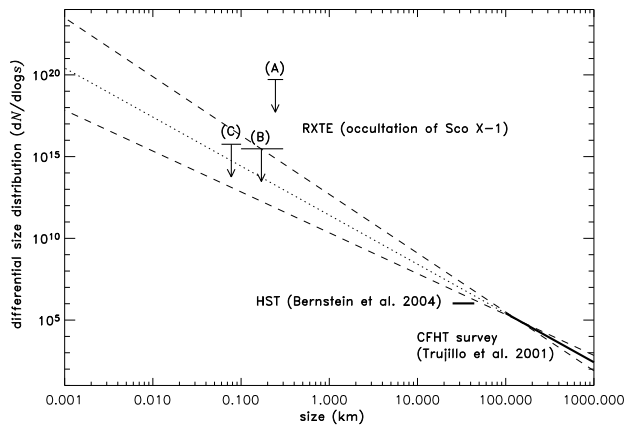}
\caption{The differential size distribution of TNOs. 
The 3 upper limits to the total number of TNOs per decade of size 
($\frac{{\rm d}N}{{\rm d}\log s}$) inferred from the 
RXTE observation of Sco X-1 are plotted as horizontal bars with
downward arrows and labels (A), (B) and (C). 
The estimation of these 3 upper limits are described in Section 4. 
Upper limit (A) is not to compare with any others in this figure, 
since all the others are for the TNO population in 30 - 50 AU, while 
upper limit (A) refers to that at about 4000 AU.
The solid and dotted lines are the best fit from the CFHT survey \citep{trujillo01}
and its extrapolation towards smaller sizes.
The dashed lines mark its 1-$\sigma$ uncertainty in the power index of the distribution.
Direct imaging of TNOs down to size of about 30 km 
were reportedly achieved by the Hubble Space Telescope 
\citep{bernstein04},
whose estimated TNO size distribution at about 30 km is a factor of 25 below the extrapolation
of the aforementiond CFHT survey. 
Optical occultation search may be able to 
detect occultation events caused by TNOs 
of kilometre size.
}
\end{figure}

The upper limits proposed above are not quite constraining,
and there are considerable uncertainties in the estimate.
Nevertheless, they point out a way to acquire information for that 
size range, which is probably impossible to achieve with optical observations.
In the visible light, the Fresnel scale at 40 AU is about 1 km.
It is difficult to detect occultation by objects much smaller than that size at 40 AU or beyond.
The detectibility is further constrained by many other issues related to the background stars and
the observing facility. 
A detailed study on the detectibility of optical occultation search can be found in
\citet{nihei07}.
It is similar for the case of X-ray occultation search.
The Fresnel scale at 40 AU for 4-keV photons is about 30 m.
Any significant occultation dip events will have a duration longer than 1 ms, 
unless the occulting body is at a shorter distance.
To detect dips in the light curves at millisecond time scales, the photon count rate needs to be high enough.
The RXTE/PCA count rate of Sco X-1 is about $10^5$ counts per second, which makes detection
of millisecond dips feasible. 
A more detailed analysis on the X-ray occultation detectibility 
with RXTE/PCA observation of Sco X-1 will be reported in another forthcoming paper.
We note here that at 4000 AU objects much smaller than 300 m will not be detected by X-ray occultation.

Most model computations expect the size distribution of TNOs in 30 - 50 AU at the small size end
falls below the extrapolation of the power law of objects larger than 100 km
(e.g.\ \citet{kenyon04,pan05,charnoz07}).  
The suggested upper limits labeled with (B) and (C)
  in Fig.\ 6 are both above the power-law extrapolation and thus
provide no discrimination power to the current theory.
To push down the upper limits to the level close to the power-law extrapolation, 
about 60 times the current data volume is needed, that is, about 4.3 Ms, 
or 50 days of continuous observation.
If a new instrument more sensitive than RXTE/PCA is available, 
other X-ray sources dimmer than Sco X-1 may also 
be used for such a search. 
Finally, if definitely true occultation events are found, 
it will have strong impact on our understanding of the early solar system.
Deployment of an array of X-ray satellites in a region of 300 m or so
will help to further affirm the reality of any detected X-ray shadows and
to yield more information about the occulting TNOs. 
\section*{Acknowledgments}

We thank Edward Morgan for his effort to organize a team to apply for new RXTE observations
of Sco X-1 in 2007 and 2008. The data used in this work is part of those observations.
Comments from Jean Swank to improve this paper are very much appreciated.  
This work was supported by the National Science Council of 
the Republic of China under grant NSC 96-2628-M-007-012-MY3.

\label{lastpage}
\end{document}